\documentclass[12pt]{iopart}

\usepackage{latexsym,epsfig, amsfonts}

\expandafter\let\csname equation*\endcsname\relax
\expandafter\let\csname endequation*\endcsname\relax
\usepackage{amsmath}
\usepackage{bm}

\begin{document}

\title{Quantum criticality of the imperfect Bose gas in $d$ dimensions }

\author{P. Jakubczyk and M. Napi\'{o}rkowski}

\address{Institute of Theoretical Physics, Faculty of Physics, University of Warsaw,
Ho\.{z}a 69, 00-681 Warsaw, Poland}

\ead{pjak@fuw.edu.pl, marnap@fuw.edu.pl }



\begin{abstract}
We study the low-temperature limit of the $d$-dimensional imperfect Bose gas. Relying on an exact analysis of the microscopic model, 
we establish the existence of a second-order quantum phase transition to a phase involving the 
Bose-Einstein condensate.  The transition is triggered by varying the 
chemical potential and persists at non-zero temperatures $T$ for $d>2$. We extract the exact phase diagram and identify the scaling regimes in the vicinity of the quantum critical 
point focusing on the 
 behavior of the correlation length $\xi$. The length $\xi$ develops an essential singularity exclusively for $d=2$. We follow the evolution of the phase diagram varying $d$. For $d>2$ our 
results agree with renormalization-group based analysis of the effective bosonic order-parameter models with the dynamical exponent $z=2$. 


\end{abstract}

\maketitle

\section{Introduction}
Exactly soluble models are rare in the theory of phase transitions and their prominent importance is hardly disputable. 
In systems displaying quantum criticality \cite{Sachdev_book, Science_spec_iss, Nature_spec_iss, PSSB_spec_iss, Belitz_05, Loehneysen_07}, 
properties specific to different spatial dimensionalities are entangled, which 
yields a rich structure of the phase diagram and leads to violations of 
standard paradigms. 
For this reason quantum-critical systems are harder to analyze, and often also more interesting, as compared to their classical counterpart. As a matter of fact, this is hardly ever the case that a 
microscopic model displaying a quantum critical point can be exactly studied both at temperature $T=0$ and for $T>0$. 

Interacting Bose systems play a distinct role in the quantum many-body physics and the last two decades brought tremendous progress in their understanding both on the experiment and theory sides. 
In particular, a number of important rigorous results concerning the Bose-Einstain condensation \cite{Lieb_book} have been established in the recent years. These are however in almost all cases 
restricted to $T=0$ and do not allow access to the unconventional universal features of the phase diagram at $T>0$ and comparison to computations based on non-exact techniques. A basic aspect of 
Bose-Einstein condensation is the actual order of the transition. In the non-interacting case it is third order (within the Ehrenfest classification), On the other hand, order-parameter based studies of 
dilute Bose systems \cite{Kolomeisky_92_1, Kolomeisky_92_2, Bijlsma_96, Andersen_99, Kolomeisky_00, Cherny_01, Crisan_02, Metikas_04, Nikolic_07} as well as related effective low-energy bosonic models for 
underlying Fermi systems \cite{Loehneysen_07, Hertz_76, Millis_93, Jakubczyk_08} 
 typically truncate 
the effective action at quartic order biasing the system towards a second-order transition. It is however well known that different fluctuation-related effects tend to change the order of both thermal 
and quantum phase 
transitions \cite{Belitz_05, Coleman_73, Halperin_74, Fucito_81, Belitz_99, Chubukov_04, Greenblatt_09, Jakubczyk_09, Jakubczyk_09_2, Yamase_11}, 
and often destabilize them towards first-order. In fact, this is hardly ever the case that the transition between the microscopic model and the effective order-parameter description can be made in a fully controlled 
way. It is also well known that Bose-Einstein condensation is sensitive to the boundary conditions and displays different properties at different dimensionalities 
$d$. For these reasons insights brought by simple interacting microscopic models soluble at $T\geq 0$ and arbitrary dimensionalities seem very valuable.   

In this paper we consider the quantum critical properties of the imperfect Bose gas \cite{Davies_72, Zagrebnov_01, Lewis_book, Berg_84} (IBG) in 
$d$ spatial dimensions. For $d=3$ and off the $T\to 0$ limit this system was rigorously studied \cite{Davies_72} decades ago using sophisticated mathematical methods. This is however only recently 
that the analysis was extended to arbitrary $d>2$ \cite{Napiorkowski_13} and it was demonstrated that in the classical regime (in the immediate vicinity of Bose-Einstein condensation at $T>0$ and 
for $d>2$) this system belongs to the 
universality class of the $d$-dimensional Berlin-Kac (spherical) model \cite{Amit_book, Gunton_67, Dantchev_book}. 
This was however so far not appreciated that the system displays a quantum critical point (QCP) and is susceptible to exact analysis also off the thermal phase transition, and even for $d\leq 2$. 

\section{Model and its solution}
The Hamiltonian of the imperfect Bose gas \cite{Davies_72} reads 
 \begin{equation}
  \hat{H}_{IBG}=\sum_{\bf k} \frac{\hbar^2{\bf k}^2}{2m}\hat{n}_{\bf k}+\frac{a}{2V}\hat{N}^2 \;, 
  \label{IBG_Hamiltonian}
 \end{equation}
using the standard notation. The mean-field interaction energy $H_{mf}=\frac{a}{2V}\hat{N}^2$ ($a>0$) emerges  
from a binary repulsive interaction $v(r)$ in the Kac limit $\lim_{\gamma\to 0}\gamma^dv(\gamma r)$, where the interaction strength is suppressed, but its 
range diverges \cite{Lewis_book}. This fact clarifies the actual physical content of the model (\ref{IBG_Hamiltonian}).  
The particles are spinless and are enclosed in a box of volume $V=L^d$; the system is subject to periodic boundary conditions. We ultimately pass to the thermodynamic limit.  
Applying a Hubbard-Stratonovich type transformation, the grand canonical partition function of the IBG can be cast in the form \cite{Napiorkowski_11}
\begin{equation}
\label{GC}
 \Xi (T,L,\mu) = -i e^{\frac{\beta\mu^2}{2a}V}\left(\frac{V}{2\pi a \beta}\right)^{1/2} \int_{\alpha\beta-i\infty}^{\alpha\beta+i\infty}dse^{-V\phi_b(s)}\;, 
\end{equation}
where 
\begin{eqnarray}
\label{phi_def}
\phi_b(s)=-\frac{s^2}{2a\beta}+\frac{\mu s}{a}-\frac{1}{\lambda^d}g_{d/2+1}(e^s)+\frac{1}{V}\ln(1-e^{s})\;.
\end{eqnarray}
Here $\beta=(k_BT)^{-1}$, $\lambda = h/\sqrt{2 \pi m k_{B} T}$ denotes the thermal wavelength, and the constant $\alpha<0$ in the integration limits is arbitrary. The Bose functions are defined via
\begin{equation}
 g_n(z) = \sum_{k=1}^{\infty}\frac{z^k}{k^n}\;.
\end{equation}
In addition to $\lambda$ and $L$ we immediately identify the lengthscales $L_T=(a\beta)^{1/d}$ and $L_\mu = (a|\mu|^{-1})^{1/d}$. We observe that $L_\mu$ and $L_T$ are not defined for the ideal 
Bose gas, where the limit $a\to 0$ is taken before the thermodynamic limit.  The other important observation is that all the scales $L_\mu$, $L_T$, $\lambda$, as well as the correlation length 
(see below), diverge at the QCP at $T=0$, $\mu=0$. 
Also note that at arbitrary $\mu$ the limit $T\to 0$ involves at least two divergent lengthscales $L_T$ and $\lambda$. One finds $L_T\gg \lambda $ for $d<2$ and $L_T\ll \lambda$ for $d>2$. 

The saddle-point approximation to Eq.~(\ref{GC}) becomes exact in the thermodynamic limit. The saddle-point equation $0=\phi_b'(s)|_{s=s_0}$ can be cast in the form 
\begin{equation} 
\label{s_point_eq}
-s_0\left(\frac{\lambda}{L_T}\right)^d+\textrm{sgn}(\mu)\left(\frac{\lambda}{L_\mu}\right)^d =g_{d/2}\left(e^{s_0}\right)+\left(\frac{\lambda}{L}\right)^d\frac{e^{s_0}}{1-e^{s_0}}\;. 
\end{equation}
Its properties strongly depend on dimensionality. In particular the Bose function $g_{d/2}(x)$ is finite for $x\to 1^-$ if $d>2$ and diverges otherwise. The solution $s_0$ to 
Eq.~(\ref{s_point_eq}) is related to the correlation length $\xi$ via \cite{Napiorkowski_11}
\begin{equation}
\label{xi_and_s_0}
 \xi = k \lambda|s_0|^{-1/2}, 
\end{equation}
where the constant $k$ depends on the boundary conditions \cite{Napiorkowski_12, Napiorkowski_13_2}, and thus $|s_0|\ll 1$ implies the divergence 
of $\xi$. The character of this divergence at the QCP is determined by $d$ and the way the limits $T\to 0$ and $\mu\to 0 $ are taken. In what follows we analyze the distinct regimes of the 
$(\mu, T$) phase diagram for 
different values of $d$. We  
extract the behavior of the correlation length from Eq.~(\ref{xi_and_s_0}). 

\subsection{$d=2$}
For $d=2$, Eq.~(\ref{s_point_eq}) always admits a finite solution $s_0<0$ in the limit $L\to\infty$. This is clear after observing that the function $g_{1}(e^{s_0})$ is monotonously increasing and 
features a logarithmic singularity for $s_0\to 0^-$ 
\begin{equation}
 g_{1}(e^{s_0}) =  -\log |s_0| + \dots \textrm{\hspace{5mm} for \hspace{0.1mm} } s_0\to 0^-.
 \label{asympt}
\end{equation} 
We consider $\lambda/L \ll 1$ and perform an asymptotic analysis of Eq.~(\ref{s_point_eq}) in the limit of vanishing $T$ and $\mu$.
If the obtained solution fulfills $|s_0|\ll 1$, we impose the 
condition $g_{1}(e^{s_0})\gg (\frac{\lambda}{L})^2\frac{e^{s_0}}{1-e^{s_0}}$ in addition to $\lambda/L\ll 1$. This assures that $L$ is large enough that the bulk scaling behavior is achieved. The solution to 
Eq.~(\ref{s_point_eq}) depends on how the limits $T\to 0$ and $\mu\to 0$ are taken. For $|s_0|\ll (L_T/L_\mu)^2$ the first term in Eq.~(\ref{s_point_eq}) is negligible. We refer to the set of values of 
$\mu$ and $T$ assuring this condition as Regime I. Obviously, in this case $\mu>0$. If we additionally require that $\lambda/L_\mu\gg 1$, we obtain $|s_0|\ll 1$ and we may use the asymptotic expansion 
Eq.~(\ref{asympt}). The other possibility $\lambda/L_\mu\ll 1$ leads to a contradiction within Regime I. For Regime I we find that achieving bulk scaling requires that $L$ is large enough that 
\begin{equation}
L/L_\mu \gg e^{\lambda^2/(2L_\mu ^2)}\;. 
\end{equation}
The consistency requirements are that
\begin{equation}
L_T/L_{\mu}\gg e^{-\lambda^2/(2L_\mu ^2)}\;, \;\;\;\; \lambda\gg L_\mu \;,\;\;\;\; \mu>0\;.
\end{equation}
This is straightforwardly translated to a relation between the thermodynamic parameters $(\mu, T)$, see Fig.~1. Under these assumptions we obtain
\begin{equation}
s_0 = -e^{-\lambda^2/L_\mu ^2}+ {\dots} =-e^{-\tau \mu/(k_BT)} + \dots 
\label{s_0_I}
\end{equation}
with $\tau=h^2/(2\pi am)$. In consequence the correlation length $\xi\sim T^{-1/2}e^{\tau\mu/(2 k_B T)}$ shows an essential singularity in the limit $T\to 0$ 
at positive $\mu$, which follows from Eq.~(\ref{xi_and_s_0}). 

We now consider the case, where $|s_0|\gg (L_T/L_\mu)^2$ and the left-hand side of Eq.~(\ref{s_point_eq}) is dominated by the first term. We refer to this case as Regime II. 
All $\mu$ and $T$ dependencies then drop out of 
Eq.~(\ref{s_point_eq}) and we find 
\begin{equation}
 s_0 = \tilde{s_0}(a,m) + ...\;,
\end{equation}
where $\tilde{s_0}(a,m)$ is a solution to the equation 
\begin{equation}
-\tau s_0 = g_1(e^{s_0})\;. 
\end{equation}
The only requirement for the system size is $\lambda/L\ll 1$ and the consistency requirement may be written as 
\begin{equation}
\beta |\mu |\ll | \tilde{s_0}(a,m) |\;.                                                                                        
 \end{equation}
It now follows from Eq.~(\ref{xi_and_s_0}) that $\xi\sim T^{-1/2}$. This agrees with the general prediction that $\xi\sim T^{-(d+z-2)/(2z)}$ when we insert the presently 
relevant value \cite{Sachdev_book} of the dynamical exponent $z=2$.

The remaining scaling regime (Regime III) occurs if $g_{1}(e^{s_0})$ is negligible as compared to both the terms on the left-hand side of Eq.~(\ref{s_point_eq}). In this case we find 
\begin{equation}
s_0 = \beta\mu \;.
\end{equation}
Consistency requires that $\mu<0$ and $-\mu\gg \beta^{-1}$. The temperature dependencies of $\xi$ cancel and we find $\xi\sim |\mu|^{-1/2}$ in agreement with the standard Landau theory. 

The $d=2$ phase diagram is summarized in Fig.~1. The crossover lines separating the three scaling regimes are straight. 
The crossover line slope is controlled by the dimensionless parameter $\tau$. The quantum-critical regime (II) expands upon decreasing $\tau$ (i.e. for strong interactions and/or large particle masses) and 
shrinks for increasing $\tau$.
The essential singularity of $\xi$ in Regime I derives from the logarithmic singularity of the Bose function $g_1$. For $d<2$ an analogous calculation yields a power-law divergence of $\xi$ 
for $\mu>0$, $T\to 0$. 

\begin{figure}[h]
\begin{center}
\includegraphics[width=8.5cm]{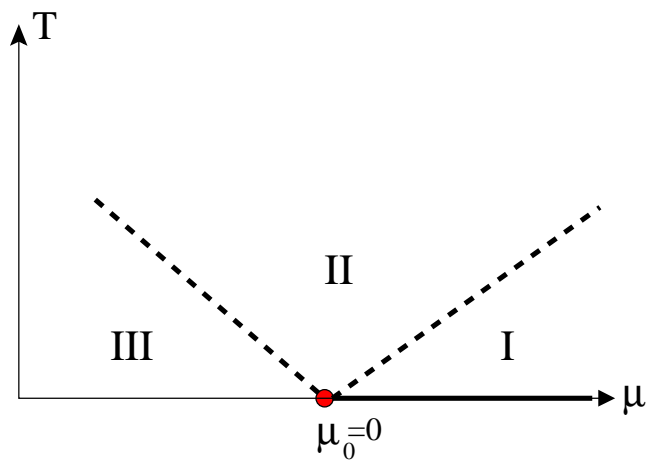}
\caption{The phase diagram of the imperfect Bose gas in $d=2$. The crossover lines separating the three scaling regimes are straight and their slope is controlled by the parameter 
$\tau = h^2/(2\pi a m)$. The correlation 
length $\xi$ in Regime I is given by Eq.~(\ref{s_0_I}), (\ref{xi_and_s_0}) and displays behavior of the type $\xi\sim T^{-1/2}e^{\tau\mu/(2 k_B T)}$. 
In Regime II we find $\xi\sim T^{-1/2}$, while in Regime III $\xi\sim |\mu|^{-1/2}$. The results in Regime II are 
consistent with the generic expectations for bosonic order-parameter models that $\xi\sim T^{-(d+z-2)/(2z)}$ when we put the dynamical exponent $z=2$. In Regime III the value $\nu=1/2$ conforms with 
Landau mean-field theory.} 
\end{center}
\end{figure}  

\subsection{$d\in \rbrack 2,4\rbrack$}
For $d>2$ the Bose function $g_{d/2}(e^{s_0})$ is bounded and in the limit $\lambda/L\to 0$ one finds a non-zero solution to Eq.~(\ref{s_point_eq}) only for $\mu<\mu_c(T)$, where 
\begin{equation}
 \mu_c(T)=a\zeta\left(\frac{d}{2}\right)\lambda^{-d}\;. 
 \label{trans_line}
\end{equation}
Here $\zeta\left(a\right)=g_{a}(1)$ denotes the Riemann zeta function. For $\mu>\mu_c$ the system is in the phase involving the Bose-Einstein condensate, where $s_0=0$ and the last term in 
Eq.~(\ref{s_point_eq}) gives a finite contribution in the limit $\lambda/L\to 0$. In this phase the system hosts long-ranged excitations and $\xi$ is infinite.
Eq.~(\ref{trans_line}) describes the shape of the line $\mu_c(T)$ of thermal phase transitions persisting down to $(\mu, T)=(0,0)$. We identify the shift exponent $\psi=2/d$ from $\mu_c\sim T^{1/\psi}$. The 
obtained value agrees with the general prediction $\psi=z/(d+z-2)$ upon inserting the dynamical exponent $z=2$. In $d=3$ we compare the obtained phase boundary Eq.~(\ref{trans_line}) to the results for the 
dilute Bose gas, calculated using leading-order perturbative expansion in the interaction coupling $u_0$ of the effective action \cite{Sachdev_book}. Somewhat surprisingly the results coincide upon the identification 
$u_0\leftrightarrow \frac{(2\pi)^3}{2}\left(\frac{\sqrt{k_B}}{h}\right)^3a$. Observe that in deriving Eq.~(\ref{trans_line}) we made no assumptions concerning the magnitude of the interactions, while 
the results of Ref.~\cite{Sachdev_book} hold only for small $u_0$ and in addition should rather refer to short-ranged microscopic interactions. 

We now perform an asymptotic analysis of Eq.~(\ref{s_point_eq}) approaching the limit $T\to 0$, 
$\mu\to 0$. Similarly to the case of $d=2$ we first consider the case of $|s_0|\ll (L_T/L_\mu)^d$, where the first term in Eq.~(\ref{s_point_eq}) is negligible. With the additional condition $|s_0|\ll 1$ we 
may expand the Bose function for $d\in \rbrack 2,4 \lbrack $ \cite{Ziff_77}
\begin{equation}
 g_{d/2}(e^{s_0}) = \zeta(\frac{d}{2} ) + \Gamma(1-\frac{d}{2})|s_0|^{\frac{d-2}{2}} + \dots\;. 
 \label{g_exp}
\end{equation}
This leads to 
\begin{equation}
|s_0| = \left[\frac{\lambda^d}{a}\left(\mu - \mu_c\right)\frac{1}{\Gamma(1-\frac{d}{2})}\right]^{\frac{2}{d-2}} + \dots\;,  
\label{s_0_dw}
\end{equation}
which applies for 
\begin{equation}
 k_BT\gg \frac{h^d}{(2\pi m)^{d/2}a}\frac{1}{|\Gamma (1-\frac{d}{2})|}\frac{\mu_c-\mu}{\mu^{(d-2)/2}} 
 \label{Ginzburg}
\end{equation}
and $\mu>0$. The result (\ref{s_0_dw}) together with Eq.~(\ref{xi_and_s_0}) lead to $\xi\sim (\mu_c-\mu)^{-1/(d-2)}$, which identifies the $\nu$ exponent in the vicinity of the thermal phase transition. 
Consistently with Ref.~\cite{Napiorkowski_13} the thermal phase transition is in the universality class of the spherical model.
The relation (\ref{Ginzburg}) defines the region displaying thermal scaling and may be understood as a Ginzburg criterion \cite{Amit_book}. In the present case, where $|s_0|\ll (L_T/L_\mu)^d$, considering $|s_0|\gg 1$ 
leads to inconsistencies. It should be appreciated that the occurrence of non-Landau critical indices at the phase transition for $T>0$ contrasts the present system to other models which are solved 
exactly by a saddle-points calculation. Examples are the reduced BCS model \cite{Muehlschlegel_61}, and pure forward-scattering models for symmetry-breaking Fermi surface deformations \cite{Yamase_05}.  

In the remaining portion of the phase diagram one identifies two scaling regimes. For $|s_0|\gg (L_T/L_\mu)^d$ we find 
\begin{equation}
|s_0| = \zeta(\frac{d}{2})\frac{a\beta}{\lambda^d}\;, 
\label{s_0_ooo}
\end{equation}
and therefore $\xi\sim T^{-d/4}$. This again agrees with the renormalization-group based prediction $\xi\sim T^{-\frac{d+z-2}{2z}}$ with $z=2$. The result (\ref{s_0_ooo}) applies for 
\begin{equation}
|\mu|\ll a\zeta(\frac{d}{2})\frac{1}{\lambda^d} \;,
\end{equation}
which defines Regime II in Fig.~2. In the opposite case ($|\mu|\gg a\zeta(\frac{d}{2})\frac{1}{\lambda^d}$) and for $\mu<0$ one finds a result analogous to Regime III in $d=2$, where $s_0=\beta\mu$.
The results obtained for $d\in (2,4)$ are summarized in the phase diagram Fig.~2.
\begin{figure}[h]
\begin{center}
\includegraphics[width=8.5cm]{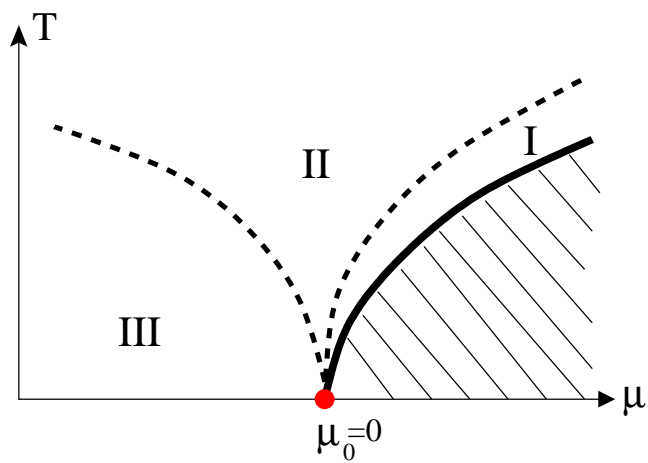}
\caption{The phase diagram of the imperfect Bose gas in $d\in \rbrack 2,4 \lbrack$. The bold line is a locus of second-order phase transitions to a low-$T$ phase involving the Bose-Einstein condensate. Its shape 
$T_c(\mu)\sim \mu^\psi$ is governed by the universal exponent $\psi=2/d$. In Regime I the system shows scaling behavior dominated by thermal fluctuations and the divergence of $\xi$ at the transition 
is governed by the critical exponent $\nu=1/(d-2)$, which is characteristic to the universality class of the spherical model. The Ginzburg line bounding Regime I from above follows the same power law as 
the $T_c$-line. The quantum-critical regime (II) occurs for $|\mu|\ll T^{d/2}$. The correlation length within this regime is characterized by the exponent $\nu=d/4$. Within Regime III the singularity 
at $(\mu,T)=(0,0)$ is cutoff by $\mu$ rather than $T$. Within this regime $\xi\sim \mu^{-1/2}$, which is typical to mean field theory and conforms with the general expectation.} 
\end{center}
\end{figure}  
We emphasize that despite the apparent triviality of the interaction term in the Hamiltonian Eq.~(\ref{IBG_Hamiltonian}), the structure of the phase diagram of Fig.~2 has very little in common with the 
one of the non-interacting gas (taking $a\to 0$ before the thermodynamic limit). Recall that the thermodynamics of the noninteracting Bose gas is defined only for $\mu\leq 0$, and condensation occurs 
exclusively for $\mu=0$ at sufficiently low temperatures. 

For $d=4$ one replaces the expansion in Eq.~(\ref{g_exp}) with $g_{d/2}(e^{s_0})=\zeta(d/2)+|s_0|\log |s_0| +...$. This leads to 
\begin{equation}
s_0 = \zeta (2) \frac{\mu_c-\mu}{\mu_c \log(\frac{\mu_c-\mu}{\mu_c})} 
\end{equation}
and yields a logarithmic correction to scaling behavior of $\xi$ in the thermal regime (I). The results obtained for $d\in (2,4)$ in the quantum critical regime (II) and in the ''$T=0$'' regime (III) apply 
also in $d=4$.

\subsection{$d>4$}
For $d>4$ and $|s_0|\ll 1$ we have 
\begin{equation}
 g_{d/2}(e^s_0) = \zeta(\frac{d}{2}) - \zeta(\frac{d}{2}-1)|s_0| + ...\;,
\end{equation}
and the problem simplifies because the first term in Eq.~(\ref{s_point_eq}) dominates over all other contributions which depend on $s_0$. One therefore obtains 
\begin{equation}
 s_0 = \beta (\mu-\mu_c) + ...
\end{equation}
in all the regimes where $|s_0|\ll 1$.

\section{Summary}
In summary, we have analyzed the low-temperature limit of the imperfect Bose gas in $d$ spatial dimensions. The model corresponds to the Kac scaling limit, where repulsive two-particle interactions are 
scaled to become progressively weaker and long-ranged. Despite superficial similarities to the ideal Bose gas, the model features a very different structure of the phase diagram, in particular in the 
properties related to 
 Bose-Einstein condensation. By an exact analysis performed at the microscopic level, we have established the existence of a quantum critical point in the phase diagram spanned by $(\mu, T)$. 
Bose-Einstein condensation persists for $T>0$ 
in dimensionalities $d>2$ and the thermal transition is in the universality class of the $d$-dimensional spherical model. The system is a rare example of a microscopic model, where one can exactly 
establish the occurrence of a second-order quantum phase transition and analyze the quantum-critical properties without a recourse to an effective model and (approximate and not always fully controllable) 
renormalization-group methods. Interestingly, the system is solved exactly 
by a saddle-point approach, but exhibits non-Landau critical indices. Instead, for $2<d<4$ and in the thermal regime, one finds the behavior characteristic to the Berlin-Kac universality class. The 
expression for finite-$T$ transition line (up to a factor defining the interaction constants) fully coincides with the result obtained from an effective order-parameter field theory at leading order in an 
expansion in interactions. This is interesting because the present calculation is valid at arbitrarily large interactions.

For $d=2$ the correlation length shows essentially singular behavior in one of the 
scaling regimes emergent in the limit $T\to 0$. In all the other cases, including $d<2$ we find power-law divergences. For $d>2$ the overall structure of the phase diagram is consistent with the predictions 
of the renormalization-group theory of effective order-parameter models with the dynamical exponent $z=2$. In particular, the vicinity of the quantum critical point splits into the classical regime, 
dominated by the thermal fluctuations, the quantum-critical regime, and the quantum regime, where the essential system properties are the same as in $T=0$. The correlation length in the phase involving the 
Bose-Einstein condensate is infinite, indicating generic coherent behavior. 


\ack
We would like to thank F. Benitez, A. Eberlein, N. Hasselmann, W. Metzner, P. Nowakowski, B. Obert, and H. Yamase for discussions and a number of very useful comments. 
We acknowledge funding by the National Science Centre via 2011/03/B/ST3/02638.

\end{document}